\documentclass[sigconf]{acmart}

\usepackage{booktabs} 
\usepackage{xspace}
\usepackage{framed}
\usepackage{mdframed}
\usepackage{csquotes}
\usepackage{xspace}
\usepackage{caption}
\usepackage{placeins}
\usepackage{subfig}

\setcopyright{rightsretained}

\acmDOI{10.475/123_4}

\acmISBN{123-4567-24-567/08/06}

\acmConference[DS+J'17]{Data Science + Journalism workshop at KDD}{August 2017}{Halifax, Nova Scotia, Canada}
\acmYear{2017}
\copyrightyear{2017}

\acmPrice{15.00}

\begin{document}

\newcommand{\ahcomment}[1]{\color{green!55!blue}{[#1 -AH]} \color{black}}
\newcommand{\bocomment}[1]{\color{blue!55!red}{[#1 -BO]} \color{black}}
\newcommand{\Rookie}{\textsc{Rookie}\xspace}
\newcommand{\snippets}{\textsc{sentence-sum}\xspace}
\newcommand{\facets}{\textsc{subjects-sum}\xspace}
\newcommand{\Q}{{\fontfamily{qcr}\textbf{Q}}\xspace}
\newcommand{\F}{{\fontfamily{qcr}\textbf{F}}\xspace}
\newcommand{\T}{{\fontfamily{qcr}\textbf{T}}\xspace}
\newcommand{\senttoserver}{$\mathbb{S}$\xspace}

\title{\Rookie: A unique approach for exploring news archives}

\author{Abram Handler}
\affiliation{%
  \institution{College of Information and Computer Sciences \\ University of Massachusetts Amherst}
  \streetaddress{140 Governors Dr.}
  \city{Amherst}
  \state{Massachusetts}
  \postcode{01003}
}
\email{ahandler@cs.umass.edu}

\author{Brendan O'Connor}
\affiliation{%
  \institution{College of Information and Computer Sciences \\ University of Massachusetts Amherst}
  \streetaddress{140 Governors Dr.}
  \city{Amherst}
  \state{Massachusetts}
  \postcode{01003}
}
\email{brenocon@cs.umass.edu}

\begin{abstract}
News archives are an invaluable primary source for placing current events in historical context. But current search engine tools do a poor job at uncovering broad themes and narratives across documents. We present \Rookie: a practical software system which uses natural language processing (NLP) to help readers, reporters and editors uncover broad stories in news archives. Unlike prior work, \Rookie's design emerged from 18 months of iterative development in consultation with editors and computational journalists. This process lead to a dramatically different approach from previous academic systems with similar goals. Our efforts offer a generalizable case study for others building real-world journalism software using NLP.
\end{abstract}

%
%



\maketitle

\begin{figure*}[!htbp]
\includegraphics[width=6.5in]{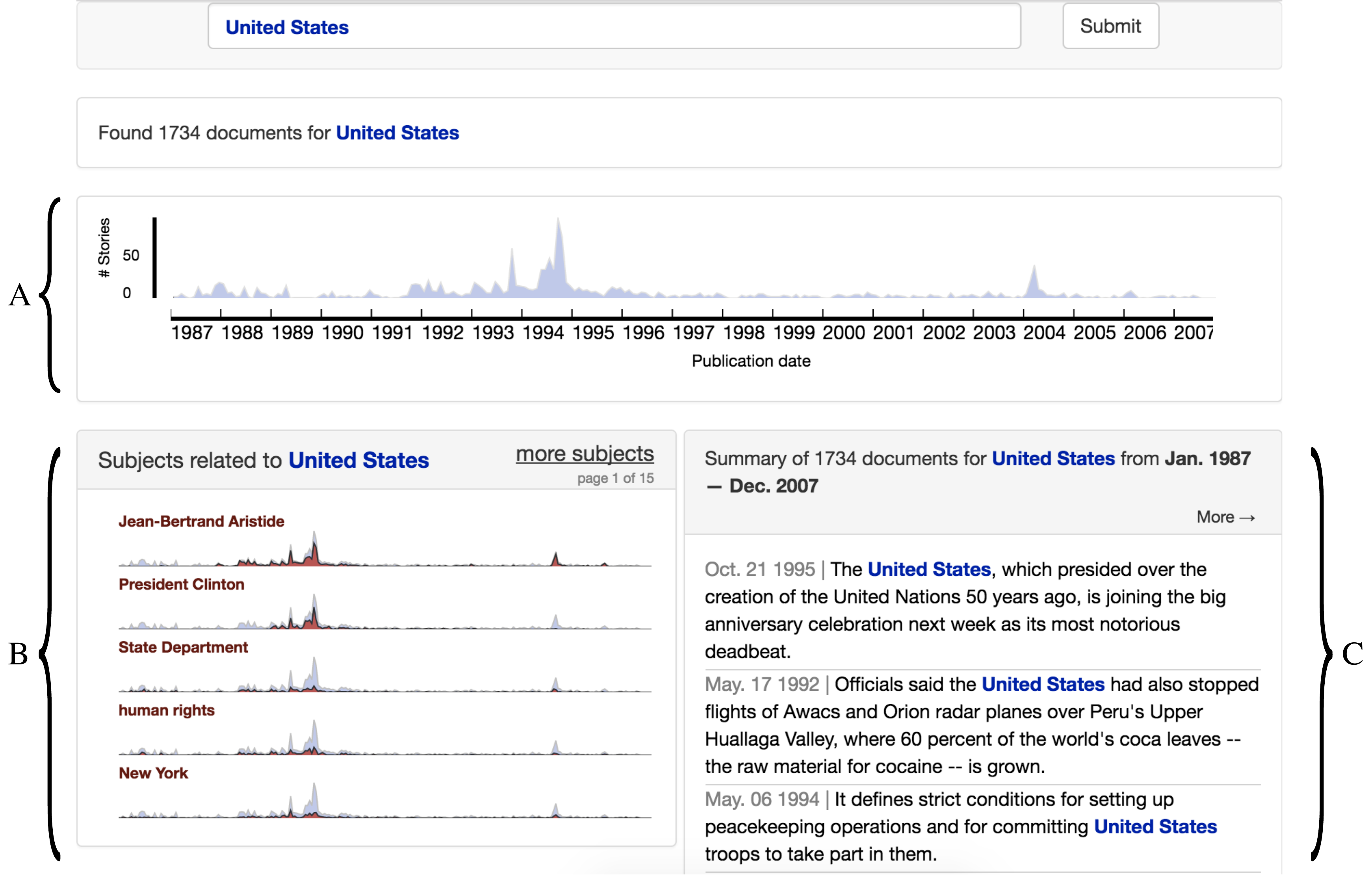}
\captionof{figure}{\label{fig:bigscreenshot} The \Rookie interface running on a corpus of \emph{New York Times} articles about Haiti.  The user has queried for ``United States.''  The interface features linked
visualization and summarization views:
(A) an interactive timeline,
(B) \facets showing automatically-generated related subjects,
and
(C) \snippets showing sentence summaries.
The temporal spikes indicate major events such as a 1994 U.S.\ intervention in Haiti and a 2004 military coup. Related subjects include specific actors in some of these events (Jean-Bertrand Aristide, President Clinton) as well as long-running topics (human rights).  Users can click and drag along the timeline to investigate specific time periods.
}
\end{figure*}

\section{Introduction}

News archives offer a rich historical record. But if a reader or journalist wants to learn about a new topic with a traditional search engine, they must enter a query and begin reading or skimming old articles one-by-one, slowly piecing together the intricate web of people, organizations, events, places, topics, concepts and social forces that make up ``the news.''

We propose \Rookie, which began as an attempt to build a useful tool for journalists. With \Rookie, a user's query generates an interactive timeline, a list of important related subjects,
 and summary of matching articles---all displayed together as a collection of interactive linked views
 (fig.~\ref{fig:bigscreenshot}). Users click and drag along the timeline to select certain date ranges, automatically regenerating the summary and subject list at interactive speed. The cumulative effect: users can fluidly investigate complex news stories as they evolve across time. Quantitative user testing shows how this system helps users better understand complex topics from documents and finish a historical sensemaking task
 37\% faster than with a traditional interface. Qualitative studies with student journalists also validate the approach.

We built the final version of \Rookie following eighteen months of iterative design and development in consultation with reporters and editors. Because the system aimed to help real-world journalists, the software which emerged from the design process is dramatically different from similar academic efforts (\S\ref{sec:related_work}). Specifically, \Rookie was forced to cope with limitations in the speed, accuracy and interpretability of current natural language processing techniques (see \S\ref{sec:iteration}). We think that understanding and designing around such limitations is vital to successfully using NLP in journalism applications; a topic which, to our knowledge, has not been explored in prior work at the intersection of two fields.

\section{The Rookie system}\label{sec:rookie}

At any given time, \Rookie's state is defined with the \textbf{user selection state},
a triple $(\Q, \F, \T)$ where:

\begin{itemize}
\itemsep0em
\item \Q is a free text query string (e.g.~``Bashar al-Assad'')
\item \F is a related subject string (e.g.~``30 years'') or is \texttt{null}
\item \T is a timespan (e.g. Mar. 2000--Sep. 2000); by default, this is set to the span of publication dates in the corpus.
\end{itemize}

\begin{figure}[!htbp]
\begin{framed}
\subfloat[A user enters \Q=``Bashar al-Assad'' in order to learn more about the Syrian civil war. ]{
  \includegraphics[width=\linewidth]{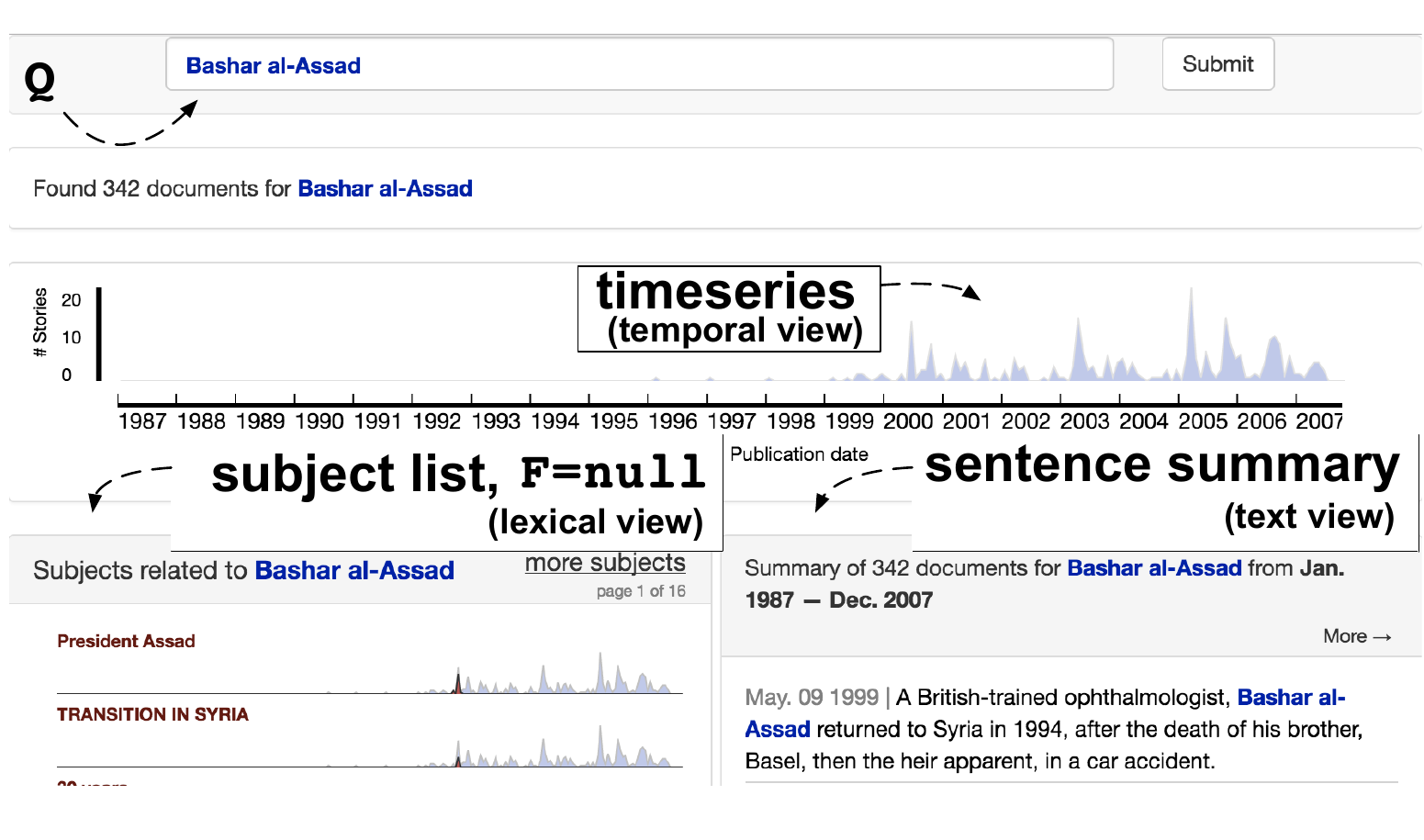}
}
\vspace{-0.05in}
\subfloat[The user zooms in to the blue spike from \T=Mar. 2000--Sep. 2000\label{fig:timeseries} to investigate. \Rookie updates \facets (fig. c) and \snippets to reflect \T. \vspace{-0.08in} ]{%
  \includegraphics[width=\linewidth]{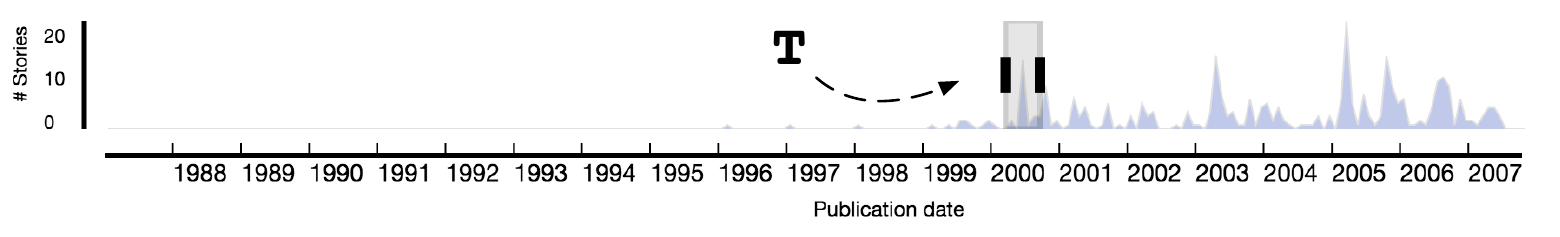}%
}

\subfloat[The user examines important subjects for \Q=``Bashar al-Assad'' during \T=Mar. 2000--Sep. 2000, displayed in \facets. The user clicks to investigate \F=``President Assad''.]{%
  \includegraphics[width=\linewidth]{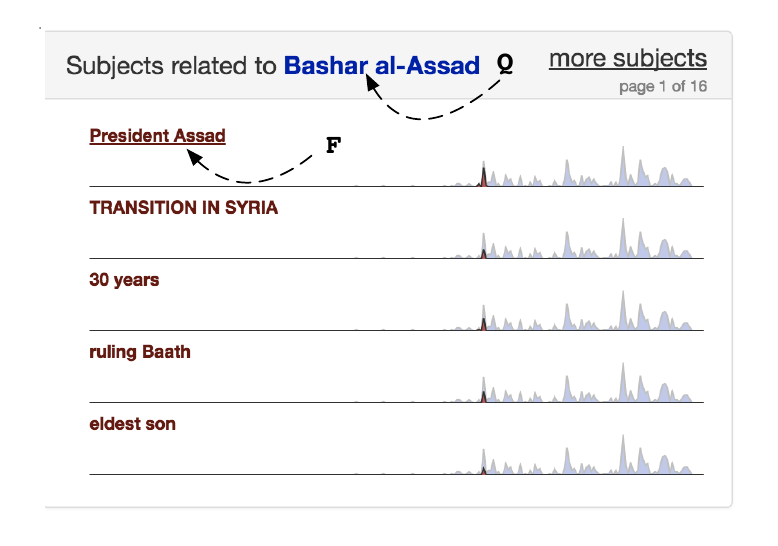}%
}

\subfloat[\Rookie now adds mentions of \F=``President Assad'' to the time series graph. \snippets updates to reflect \Q=``Bashar al-Assad'', \F=``President Assad'', \T=Mar. 2000--Sep. 2000.]{%
  \includegraphics[width=\linewidth]{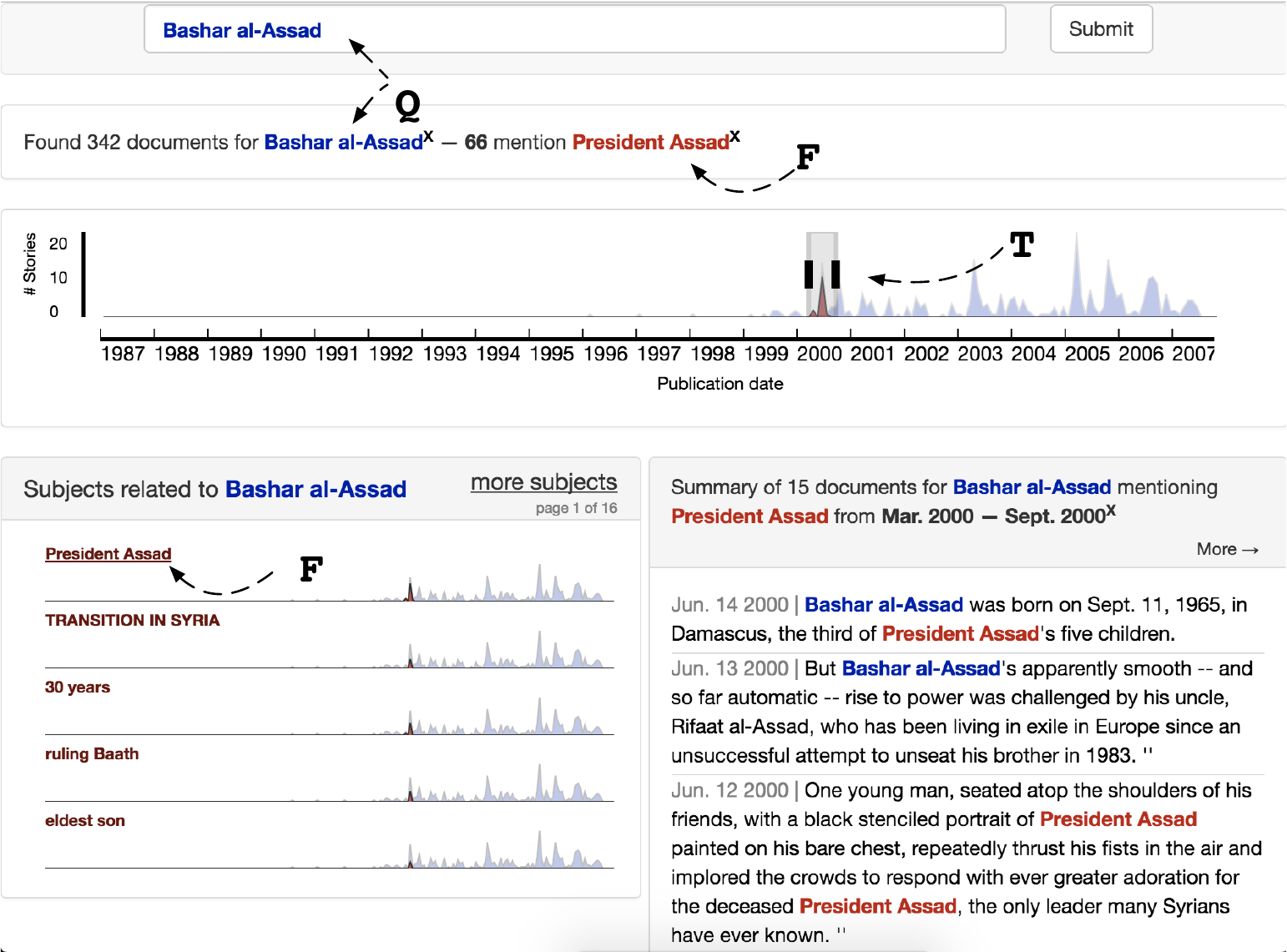}%
}

\caption{A user investigates ``Bashar al-Assad''.
}
\label{fig:interactive}
\end{framed}
\end{figure}

Users first interact with \Rookie by entering a query, \Q into a search query bar using a web browser. For example, in fig. \ref{fig:interactive}a, a user seeking to understand the roots of the Syrian civil war has entered \Q = ``Bashar al-Assad''. In response, \Rookie renders an interactive time series visualization showing the frequency of matching documents from the corpus (\S\ref{sec:timeseries}), a list of subjects in the matching documents (\S\ref{sec:facets}),
called \facets and a textual summary of those documents (\S\ref{sec:snippets}),
called \snippets.\footnote{In this example, the corpus is a collection of \emph{New York Times} world news articles from 1987 to 2007 that contain the string ``Syria''. All of the country-specific examples in this study are subsets of the same \textit{New York Times} LDC corpus \cite{SandhausNYT}.}

After entering \Q, the user might notice that ``Bashar al-Assad'' is mainly mentioned from 1999 onwards.
To investigate, they might adjust the time series slider to a spike in early mentions of Bashar al-Assad, \T=Mar. 2000--Sep. 2000 (fig.~\ref{fig:interactive}b).

When the user adjusts \T to Mar. 2000--Sep. 2000, \snippets and \facets change to reflect the new timespan (fig. \ref{fig:interactive}c). \facets now shows subjects like ``TRANSITION IN SYRIA'',\footnote{Formatting from NYT section header style.} ``President Assad'', ``eldest son'' and ``30 years''  which are important to \Q during \T. (Bashar al-Assad's father ruled for 30 years).

At this point, the user might explore further by investigating the related subject, \F=``President Assad''---clicking to select. \snippets now attempts to summarize the relationship between \Q=``Bashar al-Assad'' and \F=``President Assad'' during \T=Mar. 2000--Sep. 2000 (fig.~\ref{fig:interactive}d). For instance, \snippets now shows the sentence: ``Bashar al-Assad was born on Sept. 11, 1965, in Damascus, the third of President Assad's five children.'' If the user wants to understand this sentence in context, they can click the sentence---which opens the underlying document in a modal dialog.

\F and \Q are assigned red and blue colors throughout the interface,
allowing users to quickly scan for information. Bolding \Q and \F gives additional clarity, and helps ensure that \Rookie still works for colorblind users.


This example demonstrates how \Rookie's visualization and summarization techniques work together to offer linked views of the underlying corpus. Linked views (a.k.a.\ multiple coordinated views) interfaces are common tools for structured information \cite{Buja1991InteractiveDV, Elzen2013Multiples,OConnor2014MTE}: each view displays the same selected data in a different dimension (e.g. a geographic map of a city which also shows a histogram of housing costs when a user selects a neighborhood). In \Rookie's case, linked views display different levels of resolution. The time series visualization offers a \textbf{temporal view} of query-responsive documents, \facets displays a medium-level \textbf{lexical view} of important subjects within the documents, and \snippets displays a low-level \textbf{text view} of parts of the underlying documents. The documents themselves, available by clicking extracted sentences, offer the most detailed level of zoom.  Thus \Rookie supports the commonly advised visualization pathway: ``overview first, zoom and filter, and details on demand'' \cite{Shneiderman1996}.

Note that we use the term \textbf{summarization} to mean selecting a short text, or sequence of short texts,
to represent a body of text. By this definition, both \facets and \snippets are a form of summarization, as each offers a textual representation of the corpus---albeit at two different levels of resolution, phrases and sentences. (In the NLP literature, ``summarization'' usually means generating a sentence or paragraph length summary).

\Rookie is a web application implemented in Python.\footnote{We use the Flask (\url{http://flask.pocoo.org/}) framework with a Postgres (\url{https://www.postgresql.org/}) database and a React front end (\url{https://facebook.github.io/react}). We used the open-source search engine Whoosh (\url{https://whoosh.readthedocs.io}), which is broadly similar to Lucene, to find documents matching \Q.}

\subsection{Linked views}\label{sec:qft}

\Rookie's \textbf{user selection state} (\Q, \F, \T) picks out a set of documents $\mathbb{D}_{(\Q, \F, \T)}$,
which were published within \T, match the query, \Q in Whoosh and contain \F (if \F is not \texttt{null}). The selection state also specifies a set of sentences
$\mathbb{S}$, used to construct a summary (\S\ref{sec:snippets}). These documents and sentences are then shown to the user in the linked views, described individually in the following sections.

\subsection{\facets: Lexical view}\label{sec:facets}

\Rookie uses natural language processing methods to find and recommend a list of subjects related to
the query $\Q$, during time \T.  These subjects are presented as a concise list of terms---thus offering a \emph{lexical} view of the $\mathbb{D}_{(\Q,\T)}$ selection (fig. \ref{fig:bigscreenshot}, bottom left).

\Rookie's subject-finding algorithm works in two stages. At index time, \Rookie makes a single pass over the corpus to find and record all phrases which match certain part-of-speech patterns.\footnote{We only index subjects that occur at least five times in the corpus for use in subject list generation,
though document retrieval for $\Q$ utilizes a standard full text index.} Specifically, \Rookie uses the NPFST method from Handler \textit{et al. }\cite{handlerdennywallachoconnor} to extract phrases, which \Rookie stores in a document--phrase index. Then, at query time, \Rookie uses this index to rank phrases which occur in documents responsive to $\Q$---returning top-ranked phrases as subjects for display in the UI.



Specifically, each time a user changes $\Q$ or $\T$, \Rookie identifies all phrases which occur in the matching documents $\mathbb{D}_{(\Q, \T)}$. \Rookie then assigns each phrase a subject relevance score. Relevance scores for each subject $s$ are calculated with $\textsc{qf-idf}_s = q_s* \frac{1}{df_s}$
where the first term, $q_s$ (``query frequency''), is a count of how many times term $s$ occurs in
$\mathbb{D}_{(\Q, \T)}$ and the second term, $\frac{1}{df_s}$ (``inverse document frequency''), is the inverse of the number of documents which contain $s$ across the corpus. Highly relevant phrases occur frequently in query-matching documents, $\mathbb{D}_{(\Q,\T)}$, but infrequently overall---similar to TF-IDF and pointwise mutual information \cite{OConnor2014MTE}. \Rookie places such high-ranking phrases at the top of \facets.

Note that NP extraction often produces split or repeating phrases \cite{handlerdennywallachoconnor} such as ``King Abdullah'', ``Abdullah II'' and ``King Abdullah II''. \Rookie uses several simple hand-written string-matching rules based on character-level Levenshtein distance and token-level Jaccard similarity to avoid displaying duplicate terms. These heuristics could be improved in future work.

\subsection{\snippets: Text view}\label{sec:snippets}

\Rookie's time series visualizations offer an immediate question: what does \Q have to do with \F during \T? For example, in fig. \ref{fig:bigscreenshot}, the user might wish to learn: what does ``United States'' have to do with the phrase ``human rights'' in articles about Haiti from the early 90s?
\Rookie attempts to answer using extractive summarization---picking sentences from $\mathbb{D}_{(\Q, \F, \T)}$ that can explain the relationship.
Unlike in traditional NLP, in \Rookie the goal is not just to summarize some topic expressed by \Q (as in traditional query-focused summarization \cite{das2007survey}), but to describe what \F has to do with \Q during \T.

\Rookie also requires that any summary can be produced quickly enough to support interactive search (see \S \ref{sec:iteration}).
Thus, in building \Rookie, we found it useful to require that the client be able to generate a summary in less than half a second without server communication.\footnote{Half a second is a rough rule of thumb for acceptable latency in interactive systems, informed by work from both Nielsen \cite{Nielsen} and Liu and Heer \cite{latencyliu}.} This principle allowed us to achieve fluid, exploratory interactions. The details of \snippets are discussed below.

\subsubsection{Summary implementation: server side}\label{sec:serversum}

After \Rookie sends a user query, \Q to the server, each query-responsive document in $\mathbb{D}_{(\Q, \F, \T)}$ is permitted to send exactly one sentence to the client. This sentence is chosen by adding each sentence in the document to a two-tiered priority queue. The tier 1 score records if the sentence contains both \Q and \F (top priority), \Q or \F (medium priority) or neither \Q nor \F (low priority). The tier 2 score is simply the sentence's sequential number in the document. (Sentences that come earlier in the document get higher priority). \Rookie sends the first sentence in each document's queue to the server, along with its publication date, sentence number and tier 1 score. We use \senttoserver
to denote the set of sentences passed to the server.

\begin{figure}[!htbp]
\begin{framed}
\centering
\includegraphics[width=\columnwidth]{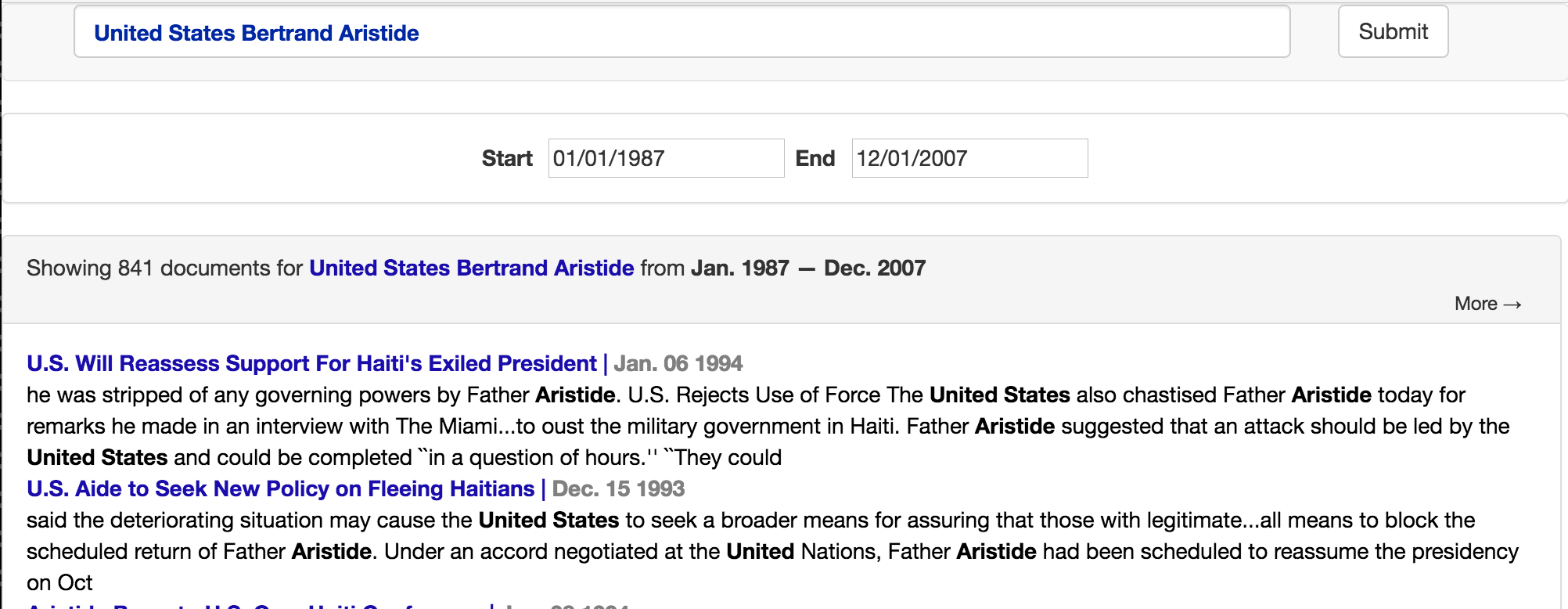}
\caption{A baseline interface ``\textsc{IR}'' interface for exploring news archives, similar to the search functionality found on many news websites.
Users enter a $(\Q,\T)$ pair query and the system returns a list of document--snippet pairs. We compare \Rookie to this interface.}
\label{fig:ir_screenshot}
\end{framed}
\end{figure}

\subsubsection{Summary implementation: client side}\label{sec:clientsum}

\Rookie seeks to help explain what \textit{happened} during a particular timespan, \T. So where traditional summarization seeks topical diversity \cite{das2007survey}, \Rookie aims for temporal diversity. It achieves this diversity by sampling sentences with probability in proportion to the count of each monthly bin. For instance, if \senttoserver contains 1000 sentences and 100 of them come from, say, March of 1993, then there ought to be a one in ten chance that a sentence from March 1993 is included in the summary.

In choosing sentences for the summary, \Rookie will first pick randomly from among sentences containing \Q and \F, then pick randomly from among sentences that contain \Q or \F and, finally, pick randomly from sentences containing from neither \Q nor \F.

\Rookie draws sentences, one-by-one, until each sentence from \senttoserver is selected and placed into a list. \Rookie allows users to page through this list (fig. \ref{fig:bigscreenshot})---starting from highest-ranked and moving towards lowest-ranked. 

\begin{figure*}[!htbp]
\begin{framed}
\centering
\includegraphics[width=\textwidth]{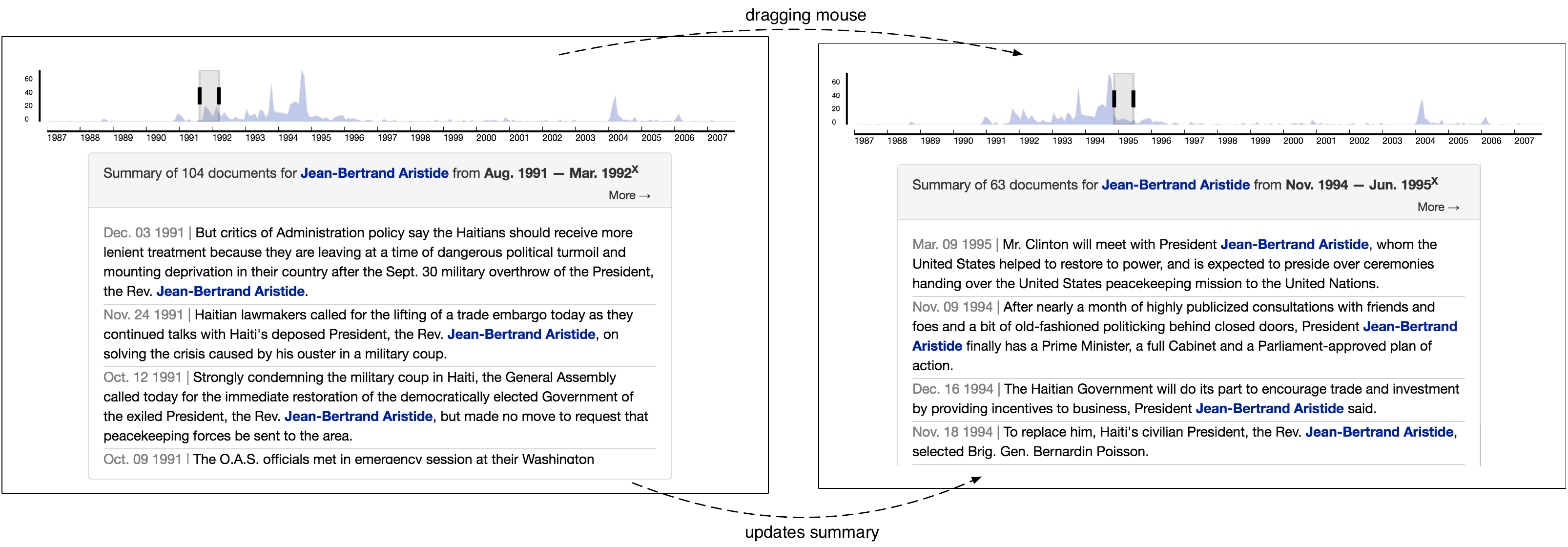}
\caption{\Rookie's summaries and time series offer linked views 
of the corpus. The \snippets panel updates in less than half a second, so users can drag a cursor across the timeline to read about unfolding events. In this example, as the user drags the rectangle to the right, they read about the 1991 coup that removed Bertrand Aristide from power---and then his return several years later with American support.}
\label{fig:drag}
\end{framed}
\end{figure*}


\subsection{Interactive time series: Temporal view}\label{sec:timeseries}

\Rookie's time series visualization is a standard line graph showing both $\mathbb{D}_{(\Q)}$ and $\mathbb{D}_{(\Q,\F)}$ across the time variable.
The y-axis represents counts of documents and the x-axis represents time. For instance, in figure \ref{fig:bigscreenshot}, the blue line shows counts of documents which match \Q=``United States''. A small copy of the time series graph for each subject is shown in the subject list
(fig.~\ref{fig:bigscreenshot}). These small graphs (``sparklines'') give cues about a subject's importance at difference times in a corpus---even if the subject is not selected.

The time series graph allows the user to specify a desired time range \T. Users can select particular areas of the time series graph by clicking and dragging a timebox \cite{Hochheiser2004DynamicQT} to create specialized summaries of certain time periods. If the user holds down their mouse, clicks the grey rectangle, and slides the mouse across the timeline, the user state changes to reflect the new \T. \facets and \snippets show the evolving relationship through time.

 \section{Evaluation}\label{sec:evaluation}

 We evaluate \Rookie using several established practices for evaluating exploratory search \cite{White2009ExploratorySB}\cite{white2008evaluating} including (1) surveys and questionnaires to measure user experience
 and (2) quantitative measurements of human performance in completing a search task.
\footnote{All studies were approved by IRB.}

 For each evaluation, we compared \Rookie to a traditional search engine,
 the baseline tool for answering any question from a collection of documents.

Traditional information retrieval (\textsc{IR})
 systems return a ranked list of document--snippet pairs in response to a user's textual query.
 Users read or skim these snippets and documents until they better understand some aspect of the corpus---possibly re-querying for new documents during the search. We implemented the \textsc{IR} baseline using Whoosh.\footnote{Like other traditional search engines, Whoosh creates small snippets which highlight portions of each query-responsive document, boldfacing matching unigrams from the query. We tuned Whoosh snippets by adjusting the \emph{top}
and \emph{surround} parameters
to help the \textsc{IR} system fairly compete with \Rookie, which displays whole sentences.
\emph{Top} controls how many ``...''-delimited fragments Whoosh returns for each document result,
while \emph{surround} is the maximum number of characters around each highlighted text fragment.
By default, Whoosh sets \emph{surround}=20, but we found that this made for choppy, confusing snippets, so we adjusted it to 50.
We then did a grid search over possible values of the \emph{top} parameter---seeking
the value that minimized the average absolute difference
in the number of characters shown for each snippet between \Rookie versus Whoosh, arriving at \emph{top}=2. All other Whoosh parameters were set to default values.}
 Because \Rookie allows limiting documents by date, we also added a frequently-downloaded datepicker widget\footnote{\url{https://www.npmjs.com/package/react-datepicker}} so that \textsc{IR} users
  can limit results to date ranges.

We compared to a traditional search engine for a number of reasons: users are familiar with Google (but not with alternatives), there are few robust implementations of text analytics research systems available and previous work rarely compares to traditional search, which we believe is a robust and powerful baseline. 

 \section{In-person group evaluation}
 \label{sec:qualitative}

While building \Rookie, we solicited ongoing feedback from working journalists. Once we were confident in the final design, we conducted a larger and more formal user test with 15 undergraduate journalism students.  Undergraduate journalism students are a good choice for a user group, as \Rookie is built for reporters and readers learning about new topics. (An additional evaluation with professional journalists would have improved our study, however, their time is very limited).

 During the study, we loaded with a corpus of 5496 articles from the \textit{New York Times} from 1987 to 2007 which mentioned the country ``Syria''. After a short tutorial demo, we presented the users with a exploratory search prompt: ``Imagine you just got your first job as a fact checker and production assistant on a world news desk. Your organization frequently covers the civil war in Syria. Use \Rookie to better understand the roots of the Syrian civil war so that you can begin contributing at your new job''.

 We gave users twenty minutes to try \Rookie using this prompt; then presented a questionnaire about their subjective experience. We did not to tell users about the design intentions behind \Rookie. We synthesize answers to each question below.

 \textbf{Q1: Did you enjoy using Rookie? What was good about it? Or bad about it?}

 Users overwhelmingly reported that they ``really enjoyed using this tool'' and found it useful ``extremely useful in doing research''. One user said: ``It made me feel like I could find things that may be buried on a more generic search engine''. Another said: ``It makes a fluid way to search through a lot of information quickly''. 

 \textbf{Q2 ``How do you think something like [Rookie] could help journalists?''}

 Many users reported that \Rookie could be helpful for journalists or other researchers starting to learn about a new topic, which was our intention in designing \Rookie. One wrote: ``As journalists, it's important to have a large-view grasp of a story before writing about it. The system could be helpful in providing both a snapshot and an ability to then dive deeper into your story''. 


 \textbf{Q3 ``When would Rookie be better than using a traditional search engine? When would it be worse?''}

 At the end of the user study, students tried researching the same topic with a traditional search engine loaded with the same corpus. We asked which tool would be better and when. Many mentioned that \Rookie would be superior if you were starting out researching a new topic---but that a traditional search engine would be superior if you already had a clear search need. As one student wrote in praising \Rookie: ``If you aren't ... familiar about the history of the topic you probably want to build some context first''.

 \section{Task completion Evaluation}

 \subsection{Historical sensemaking task}

 Gary Marchionini \cite{marchionini2006exploratory}
 distinguishes between simple fact-finding tasks and exploratory search---the latter involves activities like comprehension, interpretation and synthesis.
 These activities are difficult to measure,
 but a simple and direct way to test how well an exploratory system supports them
 is to record how long it takes a person to
 accomplish a sensemaking task which requires these behaviors. This method, sometimes called measuring ``task time,'' \cite{White2009ExploratorySB}
 can be used in conjunction with precision and recall measures \cite{Pirolli1996ScatterGatherBC}.\footnote{As White \textit{et. al} note (chapter 5), exploratory search activities are complex, so ideally
 evaluations would also measure depth of learning and understanding.
 But task time is a good place to start, as they point out.}

 We thus measured how long it takes users to correctly answer the same complex, \underline{non-factoid} research question when using \Rookie and the \textsc{IR} system. The question asks a question about the moderately complex historical relationship
 between the United States and the Haitian political figure Jean-Bertrand Aristide. We asked users to pick the correct answer from among these four: (1) The United States has been a longtime opponent of the Haitian President Jean-Bertrand Aristide.
(2) The United States has been a longtime ally of the Haitian President Jean-Bertrand Aristide. (3) The United States was initially an ally of Bertrand Aristide -- but then stopped supporting him. (4) The US government was initially an opponent of Bertrand Aristide -- but then started supporting him.

 The third answer is broadly correct, within the timespan of the corpus:
 the Clinton administration used US troops to restore Bertrand Aristide's democratically-elected government following a coup in the mid 1990s. Then, ten years later, the Bush administration did not support Aristide during a later coup.

 Users were asked to answer this question by searching for information \textit{New York Times} articles which mention ``Haiti.''  Within this corpus, articles exist describing both of these key events in this historical relationship, but there does not appear to be a complete narrative summary of this history. Users have to
 sort through, comprehend, and synthesize many
 pieces of information across multiple articles until they know the correct answer.
 The task took up to 21 minutes to complete (36 seconds minimum, 1261 maximum)
 and was fairly difficult:
 only 52\% \textsc{IR} users and 54\% \Rookie users answered correctly.

 \Rookie's evaluation simulates a practical task that a journalist might undertake in learning about a new subject: either to write an ``explainer'' piece\footnote{e.g. \url{http://www.vox.com/2015/9/14/9319293/syrian-refugees-civil-war}} or to research the historical context for current events. 

 \subsection{Experiment design}

 We employed a between-subjects design with U.S. users from
 Amazon Mechanical Turk---placing fifty workers into a \Rookie group and fifty workers into an \textsc{IR} group, and comparing their task completion times and other behaviors. Turkers had a maximum of 30 minutes to complete the task, much more than the roughly 15 minutes required.\footnote{Amazon suggests giving workers a generous maximum time limit so that they do not feel rushed} We limited our study to U.S.-based users.

 For each group, the study began with a few short screening questions---checking to make sure that workers had their volume turned on and were using a laptop or desktop (this version of \Rookie is designed for these modalities). \Rookie users also practiced interpreting a timeseries graph by explaining why mentions of Afghanistan might have spiked in the \textit{New York Times} in 2001 and 2002. Each group then watched a short video explaining their interface and task.

 During iterative prototyping (\S\ref{sec:iteration}), we observed that it takes a few minutes to learn to use \Rookie. Thus the final preliminary phase for the \Rookie group was a practice session to learn the \Rookie interface on
 a different corpus of articles mentioning ``Cuba''.
 During this tutorial, users practiced manipulating \T to select \T=1994--1995 and then answered a question about the U.S and Fidel Castro during this time period.  After this session was complete, users then were given the main task with the Haiti corpus.
 This helped ensure  the main task was measuring how long it took users to find answers using \Rookie,
 as opposed to learning how to use it. IR users did not practice using their interface.

 Users then attempted to answer the question using \Rookie or \textsc{IR}. In each case, users saw the question and answers on a panel on their screen as they completed their work. To better constrain the task, we presented each group of users with interfaces already loaded with a useful pre-filled query, rather than relying on users to think of such queries themselves. We set \Rookie to \textbf{user state}
 (\Q=``Betrand Aristide'', \F=``United States'' and \T=1987-2007) which selects 830 documents---and we loaded \textsc{IR} with the query ``Betrand Aristide United States'' which selects 841 documents. For the \textsc{IR} system, we disabled the search button during the task---users were not allowed to change the query, but could use datepickers to zoom in on certain dates.
 For \Rookie, users could not change \Q or \F, but could vary \T. Thus, this experiment measures well \Rookie's linked temporal browsing and \snippets help users learn the answer to a question---it does not measure other aspects of the full \Rookie UI.

 Users were instructed to research until they were reasonably confident in their answer, then submit a response.  They were also required to copy and paste two sentences from \snippets to support their multiple choice selection.
 By requiring evidence, we avoided junk responses and signaled to participants that answers will be scrutinized---following best practices ~\cite{Kittur:2008:CUS:1357054.1357127} for user interface studies conducted via MTurk.
 In the analysis below, we analyze time to completion in cases where workers answered the question correctly, in order to ensure we are measuring ``good faith'' \cite{Kittur:2008:CUS:1357054.1357127} attempts to complete the task.

 \subsection{Results and analysis}

 Limiting our results to the 26 workers who found the correct answer with the IR system and 27 workers who found the correct result with the \Rookie system, we find \Rookie users complete the task faster; in seconds:

 \vspace{-0.1in}
 \begin{minipage}[h]{1.7in}
 \begin{tabular}{ ccc }
  \hline
   & Mean & Std.\ Dev. \\ \hline
  Rookie & 285.1 & \footnotesize{(169.8)} \\
  \textsc{IR} & 451.0 & \footnotesize{(217.5)} \\
  \hline
 \end{tabular}
 \end{minipage}
 \begin{minipage}[h]{1.6in}
 \includegraphics[width=1.6in]{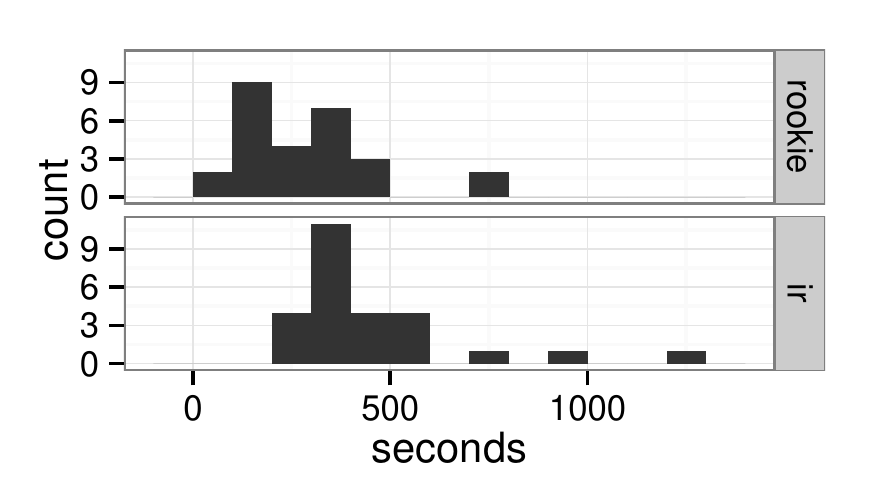}
 \end{minipage}
 \vspace{-0.01in}

 While there is considerable variation within each group,
 \Rookie users, on average, completed the task 166 seconds faster
 (using 37\% less time);
 the difference is
 statistically significant ($p=0.003$, t-test or $p=0.001$, non-parametric Mann-Whitney test).
 The fastest IR user completed the task in 210 seconds, but 12 \Rookie users
 finished faster than that.
 Interestingly, the choice of system did not affect accuracy of the results---roughly
 half of both groups got the question right,
 and among users who submitted incorrect answers, completion times were similar
  (232 {\footnotesize{(149)}} vs.\ 460 {\footnotesize{(277)}}).

 Examining system logs for \Rookie and \textsc{IR}, we find that 25 successful
 \textsc{IR} users opened individual news stories for inspection (viewing 6.4 stories on average) while only 7 successful \Rookie users opened individual stories for inspection (viewing 2.0 stories on average). (By successful, we mean answered the question correctly).
 This suggests
 that \textsc{IR} users solved the task by reading documents, but \Rookie users solved the task by reading summaries. Interestingly, the most frequently requested document by Rookie users, which describes how Aristide fled Haiti in 2004 ahead of U.S. troops, was not requested by a single IR user. The headline and Whoosh snippet for that document only describe the United States---not its relationship to Aristide. In contrast, for that document, \Rookie's \snippets shows a sentence that describes the US and Aristide.

 \section{Real-world journalism and NLP}\label{sec:iteration}

Natural language processing researchers have developed many techniques for extracting entities, relations, events, topics \cite{blei2003latent} or summaries from news archives (see Grishman \cite{Grishman2012InformationEC} and Das \cite{das2007survey} for surveys). Such work, in the words of Jonathan Stray \cite{stray}, sometimes ``discusses journalism as a potential application area without ever consulting or testing with journalists''.

\Rookie is one of a handful of projects, including Vox Civitas \cite{diakopoulos2010diamonds} and Overview \cite{Brehmer2014OverviewTD} which seek apply work from the NLP research community using feedback and input from actual readers, writers and editors.

During development, we consulted with three reporters, two news editors, a journalism professor and a new media applications developer. We spent nearly 18 months testing features and modifying the design before we began formal user testing. Several crucial themes emerged from this process. We strongly suspect that these lessons would apply to others seeking to join journalism and NLP.

 \subsection{Practical systems should handle NLP failures with grace}

 Early versions of \Rookie attempted to produce authoritative summaries of queried documents. One early version of \snippets attempted to combine multiple phrases and sentences from across documents without line breaks (similar to Zagat\footnote{e.g. http://www.zagat.com/r/franklin-barbecue-austin} reviews). Another early version showed only the top N topics in \facets, or the top N sentences in \snippets, without pagination. These designs were well-intentioned: we hoped to cut through ``information overload''~\cite{shahaf2013information} and present users with only the most important data.

 However, users reported confusion and frustration with our first attempts. They did not understand \Rookie's long and sometimes incoherent summaries---and they were annoyed when \Rookie only showed the top 5 most important sentences or phrases for display. In the first case, we had trouble accounting for discourse effects in summaries, so sentences strung together into a long summary did not necessarily make sense as a whole. In the later case, we attempted to create an authoritative, fixed-width summary (either with a list of 5 terms or a list of 5 sentences)---without allowing users to expand the summary as needed using pagination, as in the current design.

Because state of the art NLP systems are not able to produce perfectly grammatical extractive or abstractive summaries (e.g. \cite{rush2015neural}) it is thus important to design summarization interfaces which are robust to such errors. In \Rookie, we did this by (1) splitting each extracted sentence into a standalone snippet (protecting against ungrammatical or semantically nonsensical multi-sentence summaries) and (2) allowing for pagination (protecting against poor computational judgements about importance). Informal feedback improved.

This insight about summarization generalizes to other sorts of software that uses NLP judgements for computational journalism, such as newsroom software which uses NLP to find entities, extract events or resolve coreference. Designers and developers must build systems which can handle inevitable failures \cite{Nguyen2015Calib}.

 \subsection{Text visualization should support drill down to actual words}\label{sec:drilldown_viz}

 Early versions of \Rookie's time series graph simply showed the frequencies of stories through time, without the rich interactions shown in fig. \ref{fig:interactive}. However, in testing \Rookie, we found that users assumed that they could manipulate the time series graph to drill down for more detail---and were confused when they could not do so. One editor explained:
 ``It is useful to see how many stories appear in a given month. But that is not clickable. How are you helping the user by providing information that they can't act on''?

 We thus modified the time series graph to allow users to move smoothly from visualization to underlying text. Again, informal feedback improved. We think that this insight might also be applied to other text visualizations like metro maps \cite{shahaf2012trains}, entity--relation graphs \cite{Gorg2013JigsawReflections} or t-SNE plots of word distances \cite{maaten2008visualizing}. In particular, popular time series frequency visualizations such as
 the Google N-Grams Viewer\footnote{\url{https://books.google.com/ngrams/graph?content=mobile}}
 or \textit{New York Times} Chronicle\footnote{\url{http://chronicle.nytlabs.com/?keyword=mobile}},
 which show the frequency of a word or phrase through time, could be improved with some form of textual drilldown like a KWIC view or list of underlying documents. As G{\"o}rg \textit{et al.}\cite{Gorg2013JigsawReflections}
  explain following years of development of one text analytics system:
  ``interactive visualization of connections between entities and documents alone cannot replace the reading of reports''.

 \subsection{NPs, not entities (or topics)}\label{sec:related_work}

 Many text previous systems have sought to help users explore and make sense of documents \cite{wordseer2011,Gorg2013JigsawIntelligence,isaacs2014footprints, diakopoulos2010diamonds}, including software specifically designed for news archives \cite{dork2008visgets, yang2010newdle}, software specifically designed for reporters \cite{Brehmer2014OverviewTD}, and software focused on evolving topics though time \cite{ParallelTopics, Cui2011TextFlowTB, Dou2013HierarchicalTopicsVE, Wei2010TIARAAV}.

 All such systems extract some interesting "aspects" of text, and present them to users for visualization and navigation. Some find and display entities and relationships.\footnote{e.g. \url{https://www.media.mit.edu/projects/news-graph/overview/} or \url{https://neo4j.com/blog/analyzing-panama-papers-neo4j/}} Others find and display learned word clusters or ``topics'' \cite{ParallelTopics, Cui2011TextFlowTB, Dou2013HierarchicalTopicsVE, Wei2010TIARAAV}, sometimes arranged in a hierarchy. A third approach relies on manually-created tags \cite{dork2008visgets, yang2010newdle}.

 Each of these established method has limitations. State of the art NER systems \cite{FrancisLandau2016CapturingSS, Durrett2014AJM} incorrectly tag or fail to recognize entities. Learned topics can miss categories defined by domain experts, or generate topics that do not make sense \cite{Chuang2013TopicMD}. Human annotation is expensive and often infeasible.

 \Rookie thus takes a very different approach: find noun phrases (NPs) and let the user quickly browse them, drawing attention to co-occurrence relationships and phrases in context. This rapid browsing replaces topical term clustering or relation identification in other systems for exploring news text. NPs have many advantages over entities and topics.

Unlike topics, NPs can be expressed concisely and understood quickly, without reading and interpreting lists of (possibly nonsensical) words from a topic model.  Moreover, where inference for a topic model incurs latency (a major disadvantage in a user-facing system), simple lists of important NPs can be generated very quickly in response to user queries. This is discussed further in \S \ref{sec:keep_it_dumb}.

Similarly, unlike NER and relations, NPs can be extracted with very high accuracy, which is vital to user-facing systems where nonsensical output from NER systems may confuse users unfamiliar with NLP. Additionally, where NER systems require a predefined ontology, NPs work `off the shelf' on many corpora \cite{handlerdennywallachoconnor}, offering the fine-grained specificity of entity--relation systems without specialized annotations or a predefined knowledge base.

Note that unlike the NER systems that are usually more popular in computational journalism \cite[p. 4]{stray}, NP extraction is less tied to annotation decisions in labeled data and imposes fewer assumptions about the semantic types needed for an application. In \Rookie, NPs included valuable concepts like ``eye doctor'' or ``Assad family'', which are important to understand queries like $\Q$=``Bashar al-Assad'', but which do not refer to the sort of concrete entities which
 typically serve as the basis for conventional NER systems. Other work \cite{Foley2016Ranking} shows the efficacy of Rookie's NP extraction at scale.

 The particular justification and theory for \Rookie's \facets extraction method
  is discussed in Handler \textit{et al.} \cite{handlerdennywallachoconnor}, following
  an active area of NLP research in automatically identifying ``important'' phrases in corpora,
  sometimes called keyphrases \cite{chuang2012without}, multiword expressions, or facets \cite{stoica2007automating}. We consider \facets an application of faceted search. See Hearst~\cite{Hearst2009SearchUI} for more discussion.




 \subsection{Speed, correctness and interpretability are not optional}\label{sec:keep_it_dumb}

 In designing \Rookie, we required that each UI component could be generated quickly, interpreted easily by ordinary users and would never make a mistake in presenting some semantic representation of the underlying text. These requirements proved useful.

\underline{Speed}. Others \cite{heer2012interactive} have pointed out that ``To be most effective, visual analytics tools must support the fluent and flexible use of visualizations at rates resonant with the pace of human thought.'' We followed this advice in building \Rookie. In particular, NLP has developed many techniques for summarization which require several seconds of compute time \cite{mcdonald2007study}, which is much too slow for use in a UI. During the design process, we implemented one such method \cite{daume2006bayesian} using a multithreaded implementation in C which employed CVBO \cite{Asuncion2009}, a variational method for rapid inference for topic models. Our implementation still proved too slow for interactive use. The trouble is that if summarization code runs on the server, then each time a user adjusts \Q, \F or \T, \Rookie must (a) make a call across a network to fetch a new summary (b) wait for the server to generate the summary and (c) wait for the reply. Such latency costs are unacceptable in user-facing applications. Fast, approximate summarization techniques \cite{mcdonald2007study} which run client-side in the browser are an exciting possibility for future research.

\underline{Correctness and interpretability}. \Rookie's time series is considerably simpler than other text visualizations proposed for news archives, many of which show evolving themes across time \cite{ParallelTopics, Cui2011TextFlowTB, Dou2013HierarchicalTopicsVE, Wei2010TIARAAV,Havre2002ThemeRiverVT}. This was a deliberate design choice. \Rookie's line charts showing a single lexical item certainly cannot represent clusters of related vocabulary or ``topics'', nor can they show the relationships between such topics. However, accurately summarizing topical relationships and their evolution is an AI-hard research challenge. Because \Rookie was designed for real world use, we chose a visualization technique that could not confuse or mislead users by extracting and displaying nonsense clusters or missing important topics in the text. We show that Rookie's simple line charts improve user understanding, and we welcome such demonstrations for other more complex approaches.

 \section{Conclusion and future work}\label{sec:future}

 \Rookie began as an effort to develop a useful tool for news reporters. Because of this, \Rookie's design is dramatically different from earlier systems designed to help navigate archived news. Our approach offers lessons for others seeking to use techniques from NLP in the service of journalism.




\begin{acks}
Thanks to Steve Myers, Gabe Stein, Sanjay Kairam, Xiaolan Wang, Lane Harrison, John Foley, Rodrigo Zamith, Su Lin Blodgett and Katie Keith. This work was partly supported by a Knight Prototype grant to The Lens in New Orleans.\footnote{http://www.knightfoundation.org/grants/201550791/}
\end{acks}

\bibliographystyle{ACM-Reference-Format}
\bibliography{sample}

\end{document}